\documentclass[a4paper,fleqn]{cas-dc}

\usepackage[numbers,sort&compress]{natbib}

\def\tsc#1{\csdef{#1}{\textsc{\lowercase{#1}}\xspace}}
\tsc{WGM}
\tsc{QE}
\tsc{EP}
\tsc{PMS}
\tsc{BEC}
\tsc{DE}

\begin{document}
\let\WriteBookmarks\relax
\def\floatpagepagefraction{1}
\def\textpagefraction{.001}

\title [mode = title]{Rapid Recovery of Program Execution Under Power Failures for Embedded Systems with NVM}                      



\author[1]{Min jia}


\address[1]{East China Normal University School of Computer Science and Technology, China}

\author[1]{Edwin Hsing.-M. Sha^*}


%


\author[1]{Qingfeng~Zhuge}  

\author[1]{Rui~Xu}

\author[2]{Shouzhen~Gu}
\address[2]{East China Normal University Software Engineering Institute, China}


\cortext[<cor1]{Corresponding author}


\begin{abstract}
Embedded systems experience execution interruption due to power failures. 
After power is switched on, recovering the interrupted program from the initial state can cause negative impact.
Some programs are even unrecoverable.
To rapid recovery of program execution under power failures, the execution states of checkpoints are backed up by NVM under power failures for embedded systems with NVM.
However, frequent checkpoints will shorten the lifetime of the NVM and incur significant write overhead.
In this paper, the technique of checkpoint setting triggered by function calls is proposed to reduce the write on NVM.
The evaluation results show an average of 99.8$\%$ and 80.5$\%$ reduction on NVM backup size for stack backup, compared to the log-based method and step-based method. 
In order to better achieve this, we also propose pseudo-function calls to increase backup points to reduce recovery costs, and exponential incremental call-based backup methods to reduce backup costs in the loop.
To further avoid the content on NVM is cluttered and out of NVM, a method to clean the contents on the NVM that are useless for restoration is proposed.
Based on aforementioned problems and techniques, the recovery technology is proposed, and the case is used to analyze how to recover rapidly under different power failures.


\end{abstract}




{\centering {\huge Rapid Recovery of Program Execution Under Power Failures for Embedded Systems with NVM} }

{\centering 
Min Jia,  East China Normal University School of Computer Science and Technology, China

Edwin Hsing.-M. Sha,  East China Normal University School of Computer Science and Technology, China

Qingfeng Zhuge, East China Normal University School of Computer Science and Technology, China

Rui Xu, East China Normal University School of Computer Science and Technology, China

Shouzhen Gu, East China Normal University Software Engineering Institute, China

}

$Abstract$ - Embedded systems experience execution interruption due to power failures. 
After power is switched on, recovering the interrupted program from the initial state can cause negative impact.
Some programs are even unrecoverable.
To rapid recovery of program execution under power failures, the execution states of checkpoints are backed up by NVM under power failures for embedded systems with NVM.
However, frequent checkpoints will shorten the lifetime of the NVM and incur significant write overhead.
In this paper, the technique of checkpoint setting triggered by function calls is proposed to reduce the write on NVM.
The evaluation results show an average of 99.8$\%$ and 80.5$\%$ reduction on NVM backup size for stack backup, compared to the log-based method and step-based method. 
In order to better achieve this, we also propose pseudo-function calls to increase backup points to reduce recovery costs, and exponential incremental call-based backup methods to reduce backup costs in the loop.
To further avoid the content on NVM is cluttered and out of NVM, a method to clean the contents on the NVM that are useless for restoration is proposed.
Based on aforementioned problems and techniques, the recovery technology is proposed, and the case is used to analyze how to recover rapidly under different power failures.

\section{Introduction}





From smart homes to smart watches and wearable devices with monitoring have become widespread.
However, traditional batteries are not suitable for these devices due to their size and lifetime.
Therefore, how to power the currently rapidly developing wearable devices has become a challenge.
The emergence of the energy harvesting system solves the power supply problem of wearable devices, but it also brings another challenge.
The energy harvest system obtains electric energy from the surrounding environment by sensor \cite{b1}.
Electricity sources include kinetic energy, electromagnetic radiation, wind and solar energy. However, they are all unstable.
In energy harvest system, power failure may occur at any time, between any two lines of code, and within an unpredictable length of time, with almost no warning.
The programs may lose data and cannot to continue. When the power is full, the program can only restart form beginning.
Frequent power outages and charging causes resource waste and power consume. 
Worse, in some cases, large programs can never be completed because intermediate results are not saved. 
Further more, the time and energy cost are horrendous.
So, recovery is an issue for these energy harvest system.

Emerging non-volatile memory (NVM) with the advantages of low access latency, high density, byte-addressable, and non-volatile, such as phase change memory (PCM) \cite{b2}, spin-transfer-torque magnetoresistive random access memory (STT-MRAM) \cite{b3} and 3D-XPoint \cite{b4}, promises to compensate for the volatility and the small capacity of volatile memory.
Non-volatile memory technologies can help by allowing near-instantaneous recovery of in-memory state \cite{b5}.
If the applications checkpoint their state of program to nonvolatile memory before a failure, the program can be forward progress \cite{b6}. 
This technology is called a nonvolatile processor (NVP) \cite{b7,b8},  which enables the program to continue executing in the case of such an instant on/off. 
In the NVP, there is a fast NVM. 
Before each time power failure, the volatile state is copied to the NVM. 
When the power is restored the next time, the state of processor is copied back, and the program execution can be recovered. 

\begin{figure}[htb] \centering    
    			\includegraphics[width=0.56\linewidth]{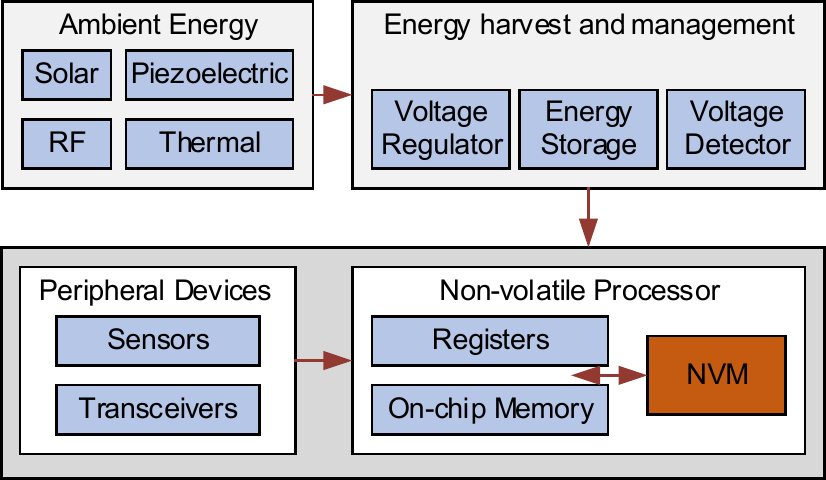}  
    		\caption{System architecture with energy harvesting system powered NVPs \cite{b9}.}     
   		\label{fig:energy harvest system}     
   	\end{figure}

Fig.~\ref{fig:energy harvest system} shows a general system architecture for NVP systems \cite{b9}.
The energy harvest system obtains electricity from the surrounding environment. Once the voltage detector detects that the voltage is low, it will back up the volatile state into non-volatile memories includes registers and volatile memory space. 
This way of backing up data is called on-demand backup. 
Embedded systems usually have a small number of registers, and their values are frequently modified during program execution \cite{b9}.
Therefore, the value of the volatile register can be designed to be automatically backed up. Registers are proposed to back up with non-volatile flop-flops (NVFFs) \cite{b7,b10,b11}.
On the other hand, the memory space was much larger than the register size. So, memory space more important to be reduced.
However, in most cases, failures occur almost without little warning. Existing studies based on voltage detector to alert and backup the state of program execution. 
They are not suitable for instant power failures.

\begin{figure}[htb] \centering    
    			\includegraphics[width=0.56\linewidth]{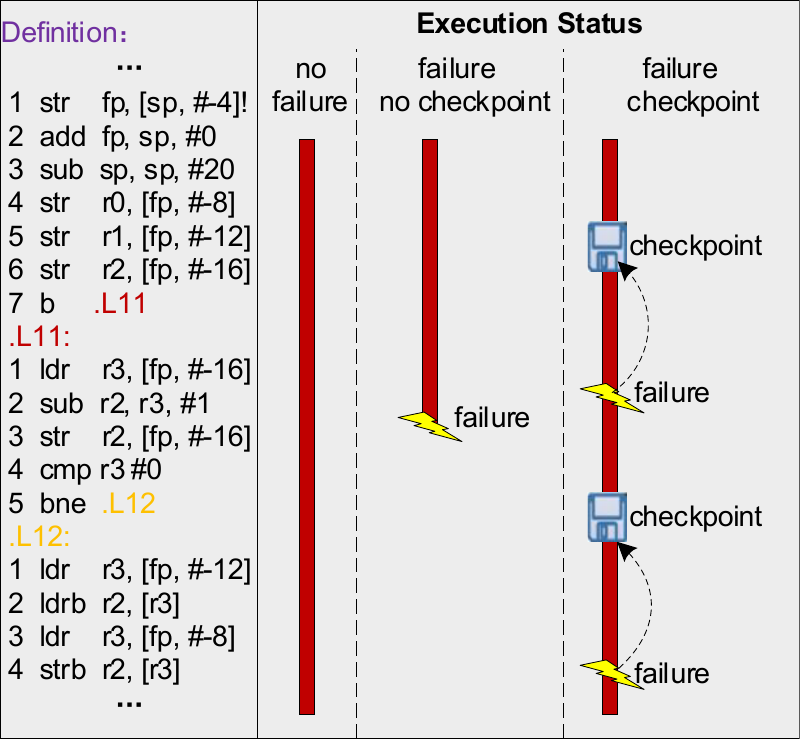}  
    		\caption{Execute the program under normal or failure or checkpoint}     
   		\label{fig:checkpoint example}     
   	\end{figure}

In another method, the processor periodically checks the processor state to the non-volatile part. Checkpointing, as shown in the Fig.~\ref{fig:checkpoint example}, saves some or all of the state of the program to NVM, without waiting for the power alarm. It allows the program to correctly continue from the checkpoint position. 
This provides the possibility to rapid recovery of program execution.
The more frequent the processor checkpoints, the fewer the rollbacks needed after recovery.
The extreme scenario is to back up every instruction, where no rollbacks are ever needed \cite{b12}.
And then after a failure, the execution of the program can be restored as quickly as possible.
Obviously, the more checkpoints are set, the faster the recovery.
Meanwhile, it will causes a lot of writing to NVM.
However, the endurance of NVM is limited and the latency and energy cost of write is high \cite{b13,b14,b15}.  
A large number of write operations would reduce the lifetime of NVM and incur significant write overhead.
Thus, it is not feasible to set a checkpoint for each instruction that has been executed.

\begin{figure}[htb] \centering    
    			\includegraphics[width=0.56\linewidth]{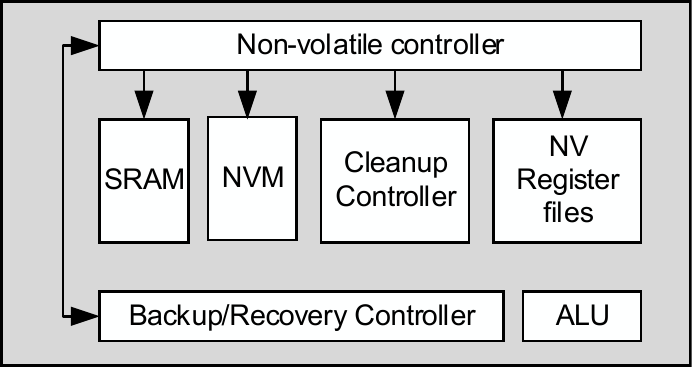}  
    		\caption{Target system architecture}     
   		\label{fig:system architecture}     
   	\end{figure}

To achieve the goal that writing NVM as less as possible while recovering quickly without voltage warning, this paper uses the $callq$ instruction to determine the checkpoints.
There are some special cases:
1) If one program without $callq$ instruction, the pseudo function call will be inserted in the program.
2) If the function call is in loop, backup is only done when the loop is executed to the power of two.
Meanwhile, the content on NVM needs to be cleaned up because there are some useless and useful information mixed on NVM.
And due to the NVM is used as on-chip memory, the size of NVM may be small, and not cleaning the useless content will cause NVM overflow.
If the clean operation is frequent, the required content may have been removed at the time of recovery.
Otherwise, there may be too much content on the NVM to hold subsequent backups.
To address this problem, a NVM clean method is proposed to clean the contents on NVM that are useless for recovery.
The detailed description for the method of clean is introduced in the subsequent paper.
In the target system, the backup operation of the state at each checkpoint is atomic.
When power failure is occurred, the backup controller will ensure that the latest state in NVM is all or nothing.
The atomicity can guarantee correct recovery.
The target system architecture is shown in Fig.~\ref{fig:system architecture}.
SRAM and NVM together serve as on-chip memory for embedded devices.
Where the NVM is used to store the program execution state of checkpoint.
After power failure and restart, the latest state of checkpoint would be load from NVM, and this program can be executed from this state rather than beginning.
Although the target system architecture of this paper is embedded systems with NVM, the design proposed is applicable to any system with NVM.

Specifically, this paper makes the following contributions:
\begin{itemize}
\item propose an universal backup and recovery system architecture, which is different from the traditional system with a power predictor and power supply from nature.
\item propose a checkpoints setting method triggered by a function call. At runtime, after receiving an counter signal and $callq$ instruction signal, the program conducts backup. 
\item propose a method to clean the contents in NVM that are useless for recovery.
\item analyze the effectiveness of the proposed scheme under different power failures. 
\item evaluate the efficacy of the proposed schemes when compared with existing checkpoint backup; analyze the NVM write sizes by the
proposed strategies.
\end{itemize}

The remainder of this paper is organized as follows. Section \ref{2} summarizes related work. Section \ref{3} provides a motivational example to illustrate the main idea of the proposed techniques. Section \ref{4} introduces the proposed scheme in detail. Section \ref{5} analyzes the backup size and the recovery speed for different cases under different power failures. Experimental results are provided in Section \ref{6}. Finally, Section \ref{7} concludes this paper.

\section{Background and Related Work}\label{2}


In this section, we first introduction related works on energy harvesting systems. 
Then, the existing work of the checkpoint scheme in the energy harvesting systems are presented. 

$Energy$ $harvesting$ $systems.$ Energy harvesting systems make the development of battery-free devices possible. Energy sources derive from solar \cite{b16}, motion \cite{b17}, radio frequency (RF), and thermal energy \cite{b18,b19} are all potential sources. 
For example, Taneja et al. \cite{b20} design a micro-solar power sensor networks, Wang et al. \cite{b11} propose a nonvolatile microprocessor powered with solar under storage-less and converter-less, Paradiso and Feldmeier \cite{b22} design a compact, wireless and self-powered pushbutton controller that functions with a piezoelectric conversion mechanism. 
However, there is an essential challenge with harvested energy. They are all unstable \cite{b23}.
Because the energy vary with its surroundings and times \cite{b24,b25,b26}.
In traditional cMOS-based processors, all the logic will be lost after shutdown and restart, causing the program to re-execute from the beginning \cite{b9}. 
In order to overcome this issue, many researches have been proposed to deploy NVM in the energy harvesting system in order to store the execution state of the program \cite{b27,b28}. 

The initial exploration into backup for energy harvesting systems focuses on flash \cite{b29,b30}. However, flash writes, beyond being energy expensive, are slow and complex.
So, another popular choice FeRAM is proposed as NVM for backup in energy harvesting systems \cite{b31,b32,b33,b34}. In comparison with Flash memory, FRAM boasts extremely low voltage writes, as low as a single volt \cite{b35}. 
Writting to FeRAM are also nearly as fast as writes to SRAM and are bit-wise programmable \cite{b36}. 
Wang et al. \cite{b11} developed a compare-and-write ferroelectric nonvolatile flip-flop for energy harvesting applications. 
Since the volatile logic is copied to the NVM, NVP can record the execution status and resume execution from the interrupted location.

However, due to the high density, high write latency and the limited endurance of NVM, reducing the size of the content to be backed up can greatly improve system performance. 
Wang et al. \cite{b37} proposed the compression-based strategies to reduce the area of NV registers. 
Compressing the data before the backup can significantly reduce the sizes of the required backup, but it also brings the overhead of compression and decompression.

$Checkpointing.$ We use the definition of checkpointing by Brenstein et al. \cite{b38}, who define it as "an activity that writes information to stable storage during normal operation in order to reduce the amount of work [the system] has to do after a failure." In the past few decades, many systems explored the checkpoints of distributed systems \cite{b39,b40,b41}. 
In order to enable the programs to be executed continuously, researchers deployed non-volatile memory in energy harvesting devices to save volatile working state at the checkpoint \cite{b9,b27,b42,b43}. 
For example, Mementos \cite{b27} is a software system for transiently powered RFID-scale devices.
It measures the code written by the user during compilation, the trigger point compares the power supply voltage with the threshold, and triggers the checkpoint to back up the volatile state to flash when the power supply voltage is less than the threshold. However, flash memory has a limited write endurance, and access to flash memory is quite slow. Xie et al. \cite{b42} proposed a instruction scheduling to reduce the backup size to NVM. 
Zhao et al. \cite{b9} proposed to reduce the NVM size of the stack backup by flexibly selecting the backup location. 
Li et al. \cite{b43} proposed a checkpoint-aware loop tiling technique. 
There are mainly two categories of checkpointing schemes: dynamic checkpointing \cite{b27} and on-demand checkpointing \cite{b9,b42,b43}.


But their technologies are all based on the system with battery warning and ignore systems that have no power alerts.
This paper solutes the aforementioned issues from the perspective of software for the system without any warning before power failure. 


\section{Motivation}\label{3}

In this section, we employ an example to illustrate the benefit of exploiting function call as checkpoint to reduce NVM size used for stack backup.

\subsection{Motivational Example}\label{3.1}
\begin{figure}[htb] \centering    
    			\includegraphics[width=0.56\linewidth]{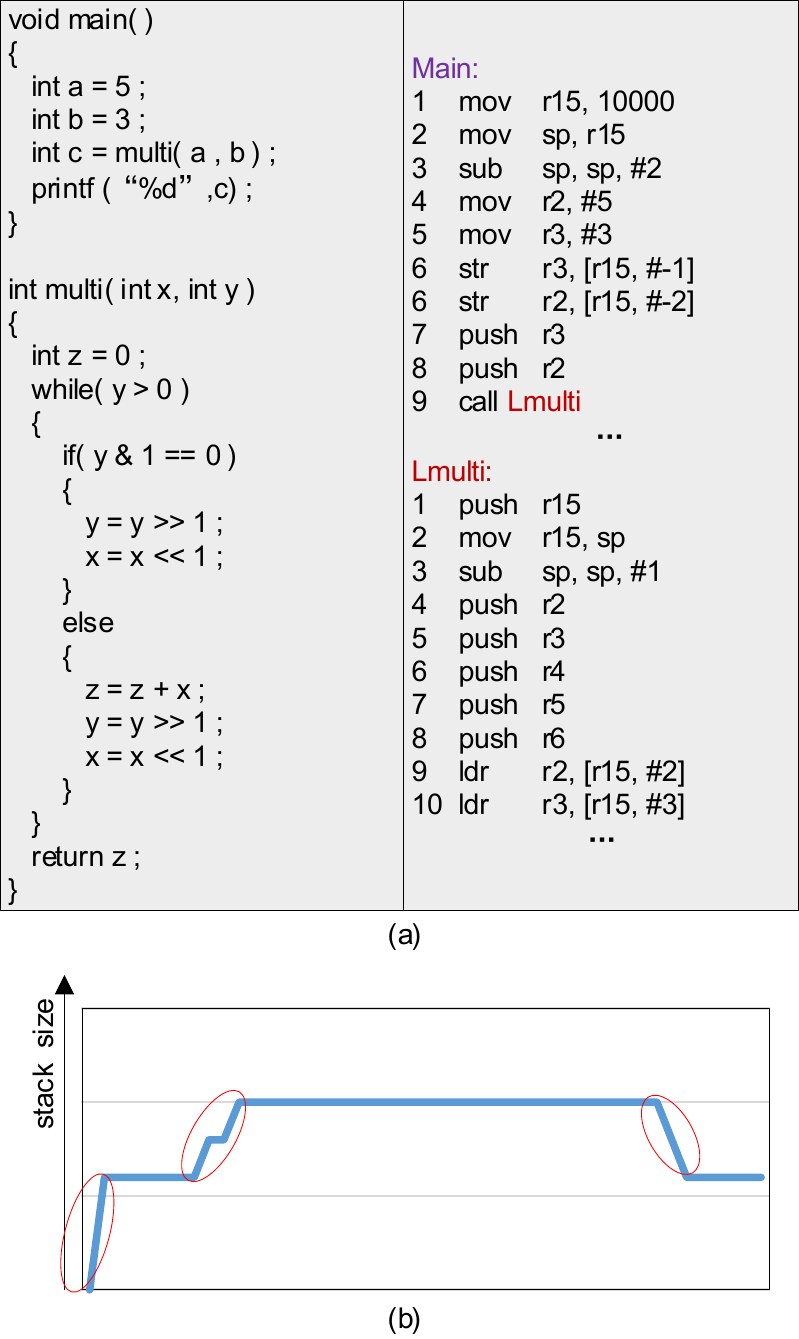}  
    		\caption{Motivation example. (a)The C program and corresponding pseudo disassembly. (b)The stack size of program execution. }     
   		\label{fig:motivation example}     
   	\end{figure}

Fig. \ref{fig:motivation example}(a) presents a sample code in the left, where the $main$ function invokes function $multi$ to get the product of two numbers. 
The corresponding pseudo disassembly instructions is shown in the Fig.~\ref{fig:motivation example}(a). 
We analyze the stack usage as shown in Fig.~\ref{fig:motivation example}(b).
At the beginning of program execution, the main function is allocated a frame to store the context information of this function. For example, local variables and temporary variables. 
It can also be seen from the disassembly instructions $sub$ $rsp$, $rsp$, $2$ that are opened up space for local variables $a$ and $b$. 
When the program enters function $multi$, $multi$ is allocated with another frame. 
In the Fig.~\ref{fig:motivation example}(b), the line in the graph shows the first trend from rising to steady, its indicate that the program into $main$ function. 
When the line in the graph shows the second trend from rising to steady, its indicate that the program into $multi$ function, and it goes inside the callee function $multi$ for execution. 
Afterwards, the line in the graph shows a trend from steady to downward, it indicates that the callee function $multi$ exits and returns to the caller function $main$ to execute.


In the case of no power detector, checkpoint backup procedures should be done in advance to make the program forward progress.
The most naive way is to use each instruction as a checkpoint \cite{b12}, which we call log-based method.
The log-based backup strategy is common used due to high reliability, because never roll backs. The main idea of the log-based backup is that each instruction is executed, all the volatile logic (including stack and registers, etc.) are written consistently to the NVM. 
Why we back up whole stack space in each time? Because, the stack size or stack content dynamically change with the execution of instructions.  
As shown at Fig.~\ref{fig:motivation example}, some instructions whose execution will affect the stack size or stack contents such as $push$, $pop$, $store$, $call$, $sub$, $return$, and so on.
In this way, it writes NVM frequently. Excessively frequent checkpoints can have high overhead.
This is not acceptable to NVM.
It is equivalent to using NVM as the main memory of the system. 

Another naive way to set the step size S, every S instructions, as a checkpoint, which we call step-based method. 
The main idea of the step-based backup is that every time S instructions are executed, all the volatile logic (including stack and registers, etc.) are written consistently to the NVM. 
Why we  back up whole stack space in each time of step size? The answer is the same as mentioned above. 
Therefore, although this method reduces the number of backups, the sizes of each backup does not change. The entire stack space needs to be backed up, and there exist a lot of redundancy.

Instead of the log-based backup method and step-based backup method, we propose to backup single stack frame at function call. 
The function call is the natural checkpoint, because if the function into callee function would openup itself stack frame, then the stack frame of caller function was fixed (here, in particular, the content is fixed).
So, when the instruction $call$ $Lmulti$ is executed, we back up the stack frame of  $main$ function. Compared with the log-base backup method, it reduces the number of backups by 46 times and the backup size of 8.2KB. Compared with the backup method of step-base, it reduces the number of backups by 4 and the backup size of 760B bytes.

\begin{figure}[htb] \centering    
    			\includegraphics[width=0.56\linewidth]{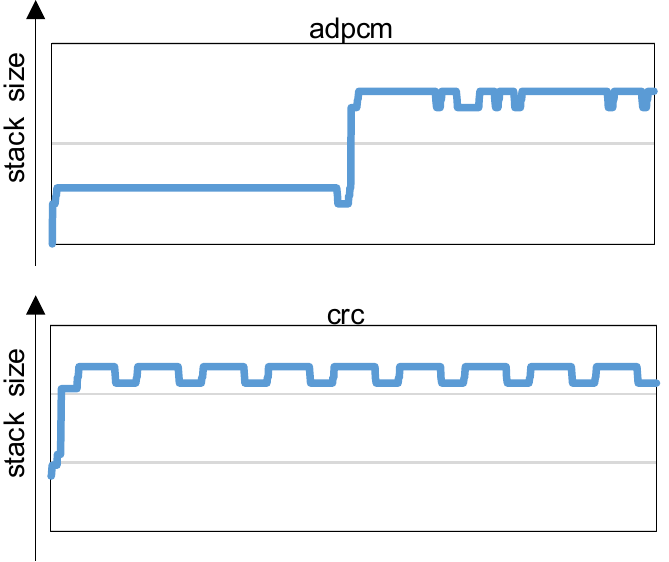}  
    		\caption{Dynamic stack sizes along program execution. Benchmarks are
from $powerstone$ \cite{b44}.  }     
   		\label{fig:satck size}     
   	\end{figure}

We presented the size of the stack space for the first 1000 instructions, as shown in Fig.\ref{fig:satck size}.
The stack behavior in the benchmark also confirms that the stack size changes dynamically as the program executes. 
As above, when the line in the graph shows a trend from rising to steady, its indicate that the function call is incurred, and it goes inside the callee function for execution. 
When the line in the graph shows a trend from steady to downward, it indicates that the callee function exits and returns to the caller function to execute. 
As a result, the system can be designed by attaching a smaller NVM. And the wear out of NVM will be reduced.
The lifetime of NVM will also increase.


\section{Design}\label{4}

\begin{figure}[htb] \centering    
    			\includegraphics[width=0.56\linewidth]{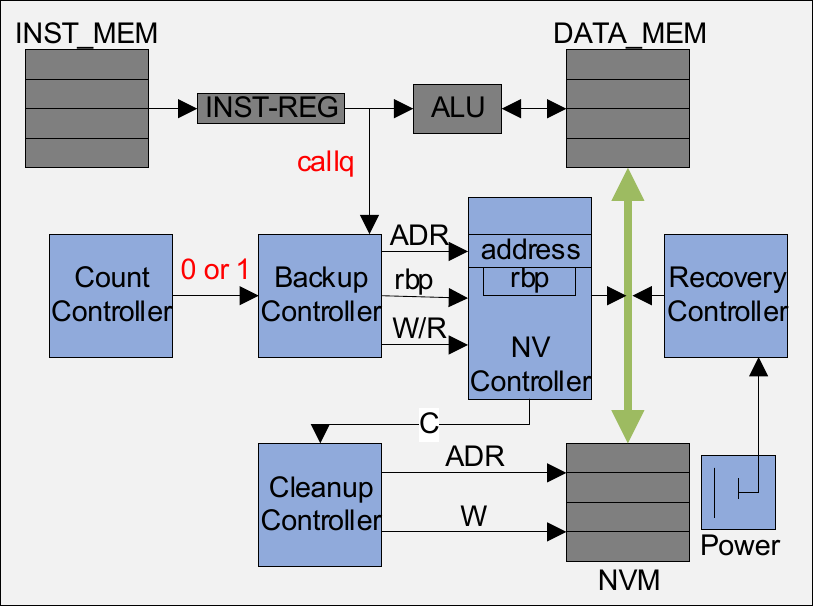}  
    		\caption{ Schematic for non-volatile processor}     
   		\label{fig:system schematic}     
   	\end{figure}

The traditional checkpoint design of system based on the energy warnings. 
When the current voltage below the voltage threshold, the voltage detector sends a warning signal to backup controller to trigger backup. 
But this design is only suitable for some systems with the power alerts, such as the systems with power provided by the natural environment (i.e. solar energy) which can be detected. 
In this paper, the schematic of non-volatile processor is shown in Fig.~\ref{fig:system schematic}. 
Different from the traditional design, the  design  proposed  is suitable for general systems under power failures without  any  warning. 
In this paper, the target system architecture is focused on embedded systems with NVM. 
The target system architecture is shown in Fig.~\ref{fig:system architecture}.
The basic idea of technology proposed is that the backup operations are triggered by counter signals and function call. 
And it also cleans up the content in NVM that is useless for recovery by comparing the $rbp$ values in different stack frames.   
The detailed process is as follows.

The schematic of non-volatile processor is shown in the Fig.~\ref{fig:system schematic}.
In the proposed scheme, the count controller sends the “0” or “1” signal to the backup controller. 
Meanwhile, the instruction register sends the instruction to backup controller. 
When “1” is issued by count controller and $callq$ instruction is issued by instruction register, the backup procedure will be triggered.
Once the backup controller is triggered, it sends the stack frame address (the address range between $rbp$ to $rsp$) and write signal to the NV controller. 
The NV controller adds the flag for the stack frame. 
The flag consists of a monotonically-increasing $epoch\_id$, validity of information, $rbp$ value and the starting and ending address for stack frame. 
The current stack frame is tagged valid and then stored to NVM, as well as the tag is stored to NVM with the corresponding stack frame. Make the checkpoint complete.
The flag will be retained in NV controller.
Then, compare the $rbp$ value of current stack frame to the $rbp$ value of past stack frame.
If the current $rbp$ is different from all $rbp$ of past stack frame, maintain current state. 
If there are a stack frame $s$ in previous stack frames that is equal to the $rbp$ of the current stack frame $s_0$, mark stack frames which $epoch\_id \geq s'epoch\_id$ and $\textless s_0'epoch\_id$ as invalid, as well as remove the corresponding tags in NV controller (Previous Stack frames are marked as invalid if and only if the current checkpoint is complete).
NV controller sends a clean signal $C$ to Cleanup controller.
Where the address that need be cleaned is included in $C$.
Once cleanup controller is triggered, it cleans the invalid information in NVM.
Due to the valid information in NVM are useful for recovery, all of the valid information are loaded from the NVM and sorted by $epoch\_id$ by Recovery controller during recovery.
Considering that registers, such as special registers and general-purpose registers, are small size and need to be backed up for each checkpoint. 
The backup of registers can be implemented according to the nonvolatile flip-flops \cite{b13}. 
Therefore, the register backup unit is not shown in the schematic of non-volatile processor. 
System performance is affected by the cost of NVM write and the time of system recovery. 
It is vital that to reduce the write cost on NVM, while rapid recovery is satisfied.

\subsection{Backup}\label{4.1}
The location of the checkpoint is very important. 
The more frequent the checkpoint, the more rapid recovery in program.
But it causes amount of cost on NVM write and reduces the life of NVM.
Thus, some methods of checkpoint setting are proposed.
To avoid power loss during a checkpointing operations, we propose a valid tail.
The valid tail writes to the end of checkpoint that includes epoch$\_$id, stack frame address, and valid flag.
A checkpoint is complete only if it contains valid tali.




\subsubsection{Call- based backup}\label{4.1.3}
\begin{figure}[htb] \centering    
    			\includegraphics[width=0.56\linewidth]{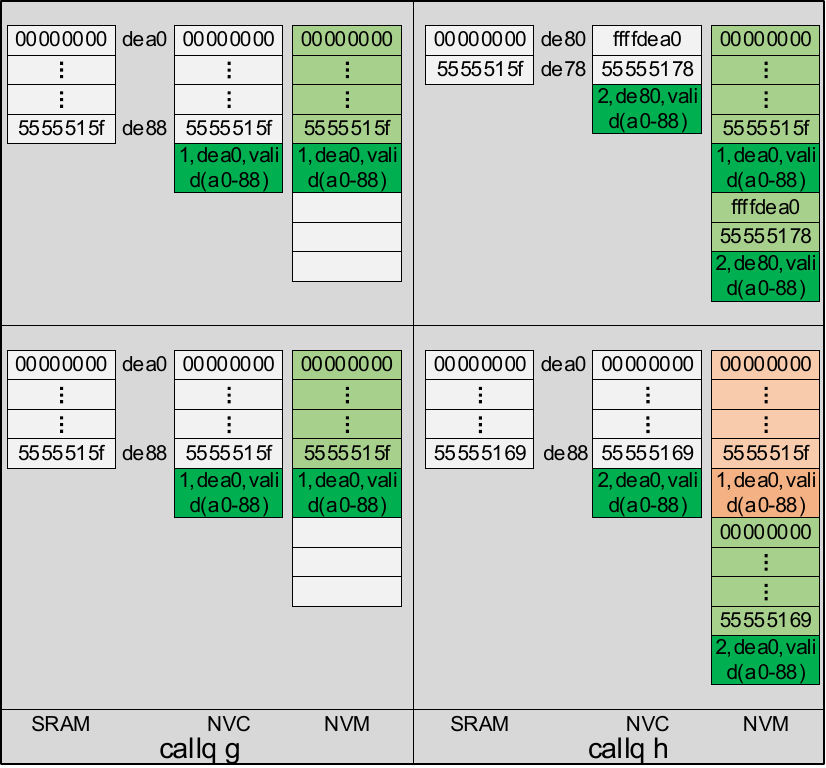}  
    		\caption{ Description of backup process  }     
   		\label{fig:detailed process in backup}     
   	\end{figure}

\begin{figure}[htb] \centering    
    			\includegraphics[width=0.9\linewidth]{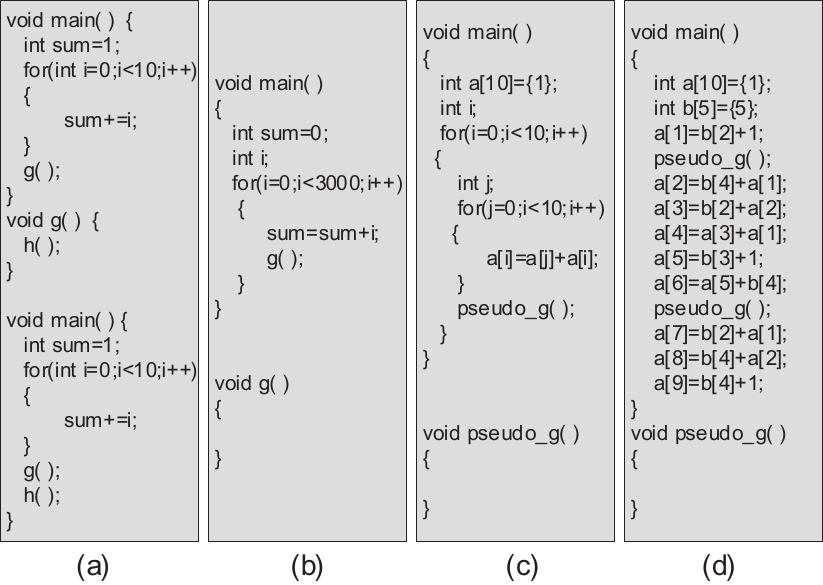}  
    		\caption{Design Example}     
   		\label{fig:design example}     
   	\end{figure}

Use the function call as the checkpoint position. 
When the program encounters a function call, it will push the address of sub-function return to the stack, and then enter the sub-function for execution. 
One program code is shown in the top of Fig.~\ref{fig:design example}(a), with multiple embedded function calls. 
When the main function calls $g$, the backup controller sends the address of stack frame of main function, the value of $rbp$ of $main$ and write signal to NV controller. 
Then, the NV controller compare the received $rbp$ and the $rbp$ in NV controller (here, the value of $rbp$ and address is empty in NV controller). 
A tag with $epoch\_id(1)$ for main function is built in NV controller.
The tag and stack frame are stored to NVM together, as shown in top left corner of Fig.~\ref{fig:detailed process in backup}.
The green content represents the tag.
And when $g$ calls $h$, the backup controller sends the address of stack frame of g function, the value of $rbp$ of $g$ and write signal to NV controller. 
The tag with $epoch\_id(2)$ for $g$ is built in NV controller.
The tag and corresponding stack frame are stored to NVM.
Then, the NV controller compare the $rbp$ of $g$ and the $rbp$ in NV controller (here, the value of $rbp$ and address in NV controller belong to the main function, the address represents the address of stack frame of main in NVM). 
The $rbp$ of $g$ and the $rbp$ in NV controller is different, so current state is maintained, the state as shown in top right corner in Fig.~\ref{fig:detailed process in backup}. 
The blue content is the newest stack frame.

Another program code is shown in the below of Fig.~\ref{fig:design example}(a). 
In this program, main calls g, main calls h. 
When the $main$ function calls $g$, the process is same as the $main$ call $g$ in preceding program.
The state of this process is shown in below left corner of Fig.~\ref{fig:detailed process in backup}.
When $main$ calls $h$, the backup controller sends the address of stack frame of main function, the value of $rbp$ of $main$ and write signal to NV controller.
A tag with $epoch\_id(2)$ for main function is built in NV controller.
Then, the NV controller compare the $rbp$ of $epoch\_id(2)$ and the $rbp$ in NV controller( here, the value of $rbp$ and address in NV controller belong to the main function, the address represents the address of stack frame of main in NVM). 
The the $rbp$ of $epoch\_id(2)$ is same as the $rbp$ of $epoch\_id(1)$ in NV controller.
The stack frame with $epoch\_id(1)$ in NVM is marked as invalid and removed the tag from NV controller.
The current state is shown in below right corner of Fig.~\ref{fig:detailed process in backup}.
The orange content represents the tag and corresponding stack frame are invalid.


In both the cases in above, all the registers stores as well as the stack frame to NVM. 
In those cases, the output signal of the counter is always 1. 
At the same time, the register unit also needs to be backed up. Because in this paper, an X86-64 machine is used, so the registers include sixteen 64bit general-purpose registers, six 16bit segment registers, one 64bit eflag register and one 64bit instruction pointer register(RIP), totaling 156B. 
For both of the programs as shown in Fig.~\ref{fig:design example}(a), when executes the "$callq\ g$" instruction, total 188B content are backed up. 

\subsubsection{Exponential incremental call-based backup}\label{4.1.4}
If the function call in a loop, as shown in Fig.~\ref{fig:design example}(b), each function call in the loop triggers a backup mechanism, which will cause multiple writes to the NVM. 
Thus, an exponential incremental call-based backup approach is proposed. 
We assume the base is 2. 
That is, the backup controller can be triggered at num =1, 2, 4, 8, 16... and the last loop (the “num” represents loop time). 
The counter will output 1 only when the loop is executed to the power of 2, otherwise it will output 0. 
Therefore, when it is not a power of 2, even if the $callq$ instruction is issued, the backup controller is disable to triggered. 
When it is a power of 2, and the $callq$ instruction is issued, the backup controller is able to triggered based on the call-based backup method policy.  
In this way, the program code as shown in Fig.~\ref{fig:design example}(b) backs up a total of 3.25KB. 
If apply the call-based backup method in every cycle, will be backed up 750KB. 
It is obviously that the exponential incremental call-based backup method can reduce the backup content by nearly 230x.
The backup operations and content are same as call-backup method.

\subsubsection{Pseudo call-based backup}\label{4.1.5}
If the program has not a function call, or a large number of instructions are executed still does not encounter a function call, the pseudo-function call will be inserted to as the checkpoint position before the program compilation. 
The basic idea follows the call-based backup and the exponential incremental call-based backup, the detailed backup mechanisms are described as the part of call-based backup method. 
The inserts mechanism based on the follow. 
If there is only one layer of loop and the number of loops is relatively small, then the pseudo function is inserted outside the loop. 
This case is similar to the main function of the code block shown in Fig.~\ref{fig:design example}(b), and the g function is equivalent to the inserted pseudo function call. 
If there is only one layer of loop and the number of loops is relatively large, then the pseudo function is inserted inside the loop. 
This case is similar to the main function of the code block shown in Fig.~\ref{fig:design example}(b), and the $g$ function is equivalent to the inserted pseudo function call. 
If there are multiple layers of loops, it is added inside the outermost loop. 
This method is shown in Fig.~\ref{fig:design example}(c). 
In this case, our method backs up 1.5B. 
If the program has not a function call, or a large number of continuous instructions does not exist a function call, we set a threshold (T), and insert a pseudo function call when the threshold is reached. 
For the code shown in Fig.~\ref{fig:design example}(d), we assume that T is 20 in this example. 
In practice, the value of T can be designed according to actual needs. 
Therefore, under this setting, our method needs to back up 640B.

The foregoing shows that the proposed approaches to reducing write NVM is effective.

\subsection{NVM Cleaning}\label{4.2}
The NVM cleanup is very important. 
Because, there are many invalid information in NVM and NVM overflow would incur if never cleanup based on backup technology proposed.
So if there are many data chunk on NVM (some are valid, and some are invalid), the invalid blocks not only occupy the space of NVM, but also affect the performance of recovery. 
So, we should clean the invalid chunk in NVM. 
 We ignore the write operations caused by NVM cleanup. 

The schematic of the NVM cleaning process is as shown in Fig.~\ref{fig:system schematic}. 
In the proposed scheme, the cleanup procedure is triggered by the clean signal (C). 
The clean signal includes invalid address of stack frame and write signal. 
Upon the clean signal is issued by NV controller, the cleanup controller was triggered to clean up the invalid stack frame in NVM. 
We analysis clean process on the program as shown in Fig.~\ref{fig:design example}(a). 
When $main$ calls $h$, the $rbp$ of $epoch\_id(2)$ is same as the $rbp$ in NV controller (because the rbp in NVC belongs to main function). 
Stack frame of $epoch\_id(1)$ is marked as invalid, as well as remove the corresponding tag in NV controller.
NV controller sends a clean signal to Cleanup controller.
The cleanup controller is triggered to clean up the invalid address of main stack frame with $epoch\_id(1)$, and then, the address is released.

\subsection{Recovery}\label{4.3}
Depending on the backup policy and NVM cleanup, the content on NVM can ensure the recovery is correctly.
The Fig.~\ref{fig:system schematic} shows the  recovery schematic. 
When the power is restored, it will send a signal to the recovery controller.
The recovery controller is triggered and sends the read signal to copy the data from the NVM to SRAM.
In this procedure, the recovery controller reads the all stack frame on the NVM.
They are valid due to the cleanup mechanism.
Then sorting by the marked $epoch\_id$ of each stack frame to get the correct stack space when recovery. 
If the failure occurs in the cleanup process, the recovery controller reads the all stack frame on the NVM during recovery, but some are invalid. 
So, the recovery controller not only sorts the stack frame by the $epoch\_id$, but also ignores the invalid stack frame. 
The tag of invalid stack frame will be send to NV controller by recovery controller to execute cleanup operation.
The recovery of the register is implemented by nonvolatile flip-flops [10].

\section{Case Studies}\label{5}




In this section, an example is used to analyze the backup and recovery under different power failures.
The failures would occur in backup process, cleanup process and program execution process, respectively. 


 \begin{figure*}[htbp] \centering    
    			\includegraphics[width=0.99\linewidth]{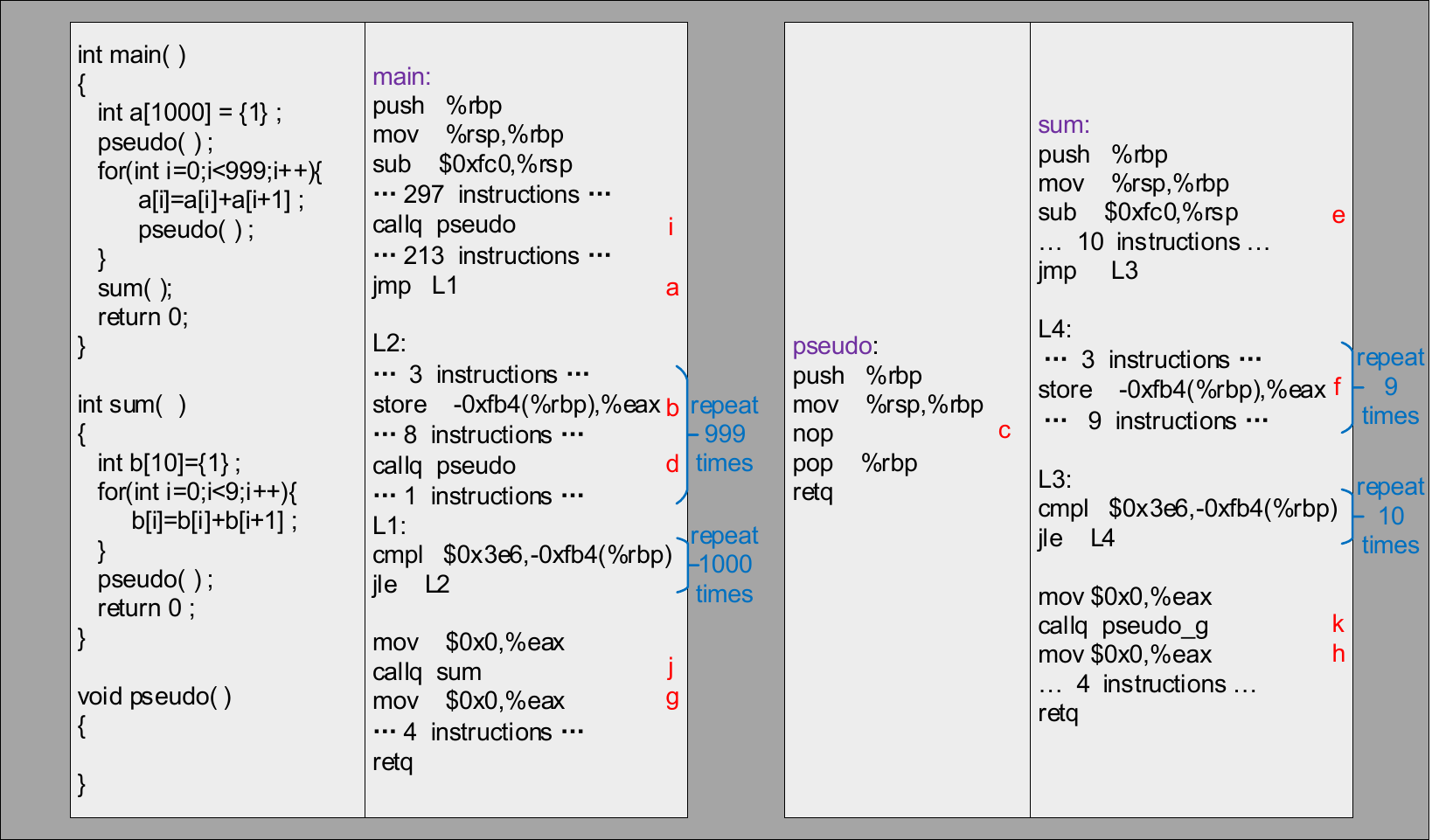}  
    		\caption{The example of case study.}     
   		\label{fig:case study}     
 \end{figure*}

We analysis the C program and its corresponding pseudo assembly code are shown in Fig.~\ref{fig:case study}. 

It includes all the methods mentioned in the section \ref{4.1} part( we set the T=300, S=10). 
We analyze the performance of the call-based backup method and exponential incremental call-backup method.

\subsection{Program execution process}
By default, the failures occurs in program execution process, the backup process and cleanup process are done normally and correctly. 
We assume that in the following analysis, failures of program execution process all occur when a certain instruction is about to be executed. Because the failure during the execution of the instruction is equivalent to the instruction not being executed, the failure after the execution of the instruction is equivalent to the failure when the next instruction is about to be executed.
In this process, we also compare with log-backup under different failures cases.

\subsubsection{Case1}
The failure occurred at the location identified by $i$, and both the call-based backup method and exponential incremental call-backup method can recover from the "$push \ \%rbp$" instruction in the $pseudo$ function. 
In this case, 4.2KB contents are backed up and 218 instructions are rolled back. 
But for the log-based backup method, there are 1231.64KB contents to written NVM. The step-based backup method, there are 123.6KB contents needed to backup and 10 instructions to roll back.

\subsubsection{Case2}
The failure occurred at the location identified by $b$ in the 500th loop. 
For the call-based backup method, the latest checkpoint is the instruction of "$callq \ pseudo$" in 499th loop, and then the program restart from "$push \ \%rbp$" in 499th call $pseudo$. 
There are only 11 instructions to rollback, but 2MB data to store. 
For the method of exponential incremental call-based backup, it needs to be executed from the 256th call to the $pseudo$ function, causing 5114 instruction rollbacks. 
But compared to the call-based backup method, the exponential incremental call-based backup method only needs to back up 41.7KB. 
The log-based backup method needs to backup 40.05MB to ensure no rollback.
The step-based backup method, there are 2253.9KB contents needed to backup and 18 instructions to roll back.

\subsubsection{Case3}
The failure occurred at the location identified by $c$ in the 128th loop in the main function. 
Both the call-based backup method and the exponential incremental call-based backup method can start execution of the "$push \ \%rbp$" instruction in the $pseudog$ function and only causes 2 instructions to rollback. 
But, compared the call-based backup method to exponential incremental call-based backup method, the exponential incremental call-based backup method is better, because only needs to store 37.55KB. 
But, the call-based backup method needs to write 535.17KB. 
However, the log-based backup method backs up 11.81MB to ensure no rollback. The step-based backup method, there are 733.48KB contents needed to backup and 12 instructions to roll back.


\subsubsection{Case4}
The failure occurred at the location identified by $d$ in the 513th loop in the main function. 
Both the call-based backup method and exponential incremental call-backup, the program needs to be executed from "$push \ \%rbp$" instruction in 512th $pseudo$ called. Because the 512th call to the $pseudo$ function is the checkpoint for these method and only caused the rollback of 20 instructions. in this case, the call-based backup method needs to back up 2.09MB, while the exponential incremental call-based backup method only needs to back up 49KB. However, the log-based backup method backs up 41.09MB to ensure no rollback. The step-based backup method, there are 2311.08KB contents needed to backup and 9 instructions to roll back.

\subsubsection{Case5}
The failure occurred at the location identified by $e$.
Both the call-based backup method and the incremental call-based backup method, it is necessary to start execution from the "$push \ \%rbp$" instruction of the sum function, which only causes the rollback of 3 instructions. 
The incremental call-based backup method only needs to back up 50.06KB, while the call-based backup method needs to back up 4.07MB. 
However, the log-based backup method backs up 78.11MB to ensure no rollback. The step-based backup method, there are 8.2MB contents needed to backup and 7 instructions to roll back.

\subsubsection{Case6}
The failure occurred at the location identified by $f$ in the 5th loop in the $sum$ function. 
Both the call-based backup method and the incremental call-based backup method, it is necessary to start execution from the "$push \ \%rbp$" instruction of the sum function. 
Both methods caused a rollback of 79 instructions, but the exponential incremental call-based backup method only needs to back up 50.06KB, while the call-based backup method needs to back up 4.07MB bytes of content. 
However, the log-based backup method backs up 78.42MB to ensure no rollback. The step-based backup method, there are 8.22MB contents needed to backup and 15 instructions to roll back.

\subsubsection{Case7}
The failure occurred at the location identified by $h$. 
Both call-based backup method and the exponential incremental call-based backup method, can start execution the $push \ \%rbp$" instruction of the $pseudo$ in $sum$ function, causing 5 instructions to roll back. 
The call-based backup method needs to back up 4.08MB, while the exponential incremental call-based backup method only needs to back up 59.29KB. 
However, the log-based backup method backs up 78.74MB to ensure no rollback. The step-based backup method, there are 8.25MB contents needed to backup and 8 instructions to roll back.

\subsubsection{Case8}
The failure occurred at the location identified by $g$. 
Both call-based backup method and the exponential incremental call-based backup method, can start execution the $push \ \%rbp$" instruction of the $pseudo$ in $sum$ function, causing 11 instructions to roll back. 
The call-based backup method needs to back up 4.08MB, while the exponential incremental call-based backup method only needs to back up 59.29KB. 
However, the log-based backup method backs up 78.77MB to ensure no rollback. 
The step-based backup method, there are 8.25MB contents needed to backup and 6 instructions to roll back.

\begin{figure}[htb] \centering    
    			\includegraphics[width=0.56\linewidth]{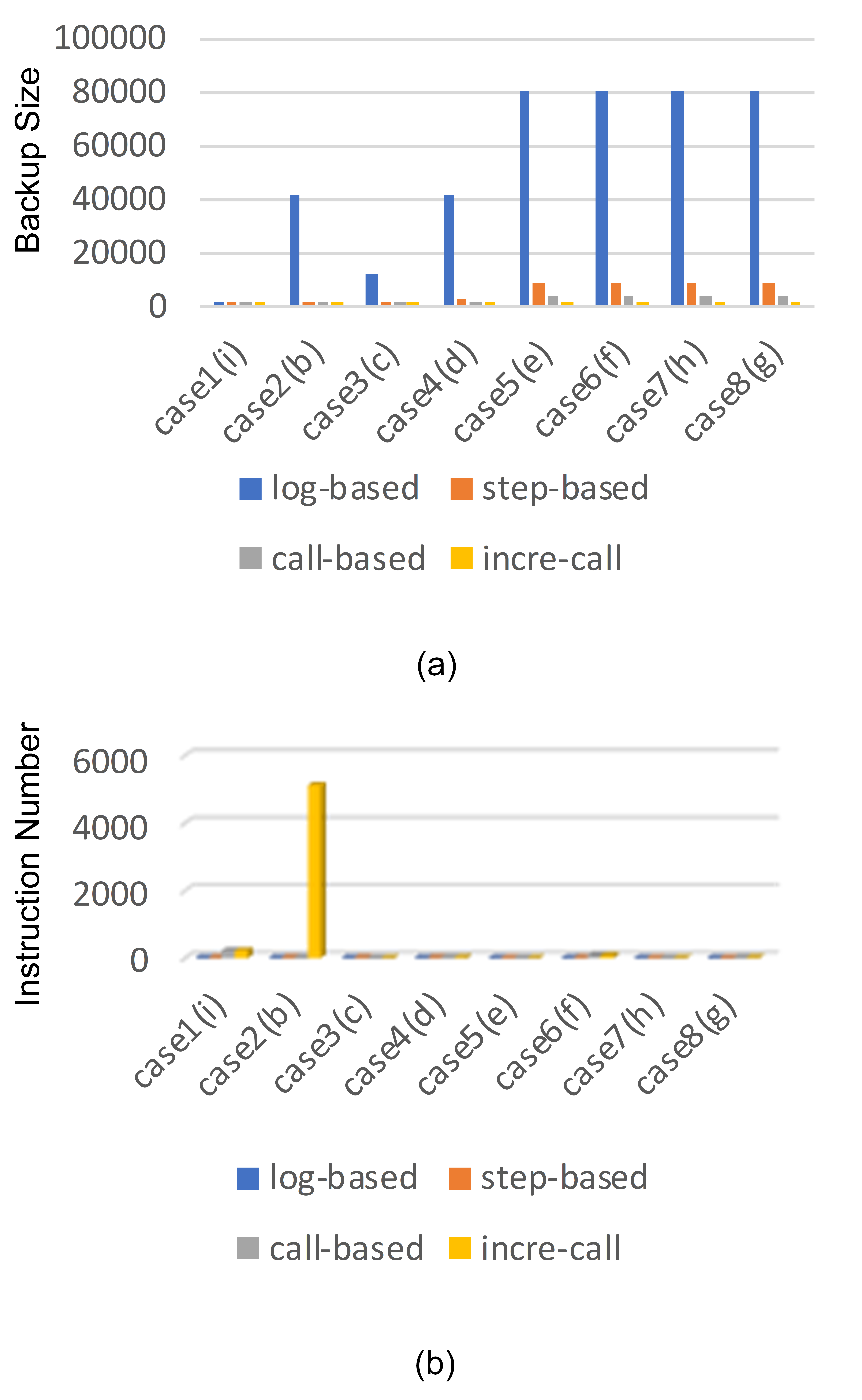}  
    		\caption{The case studies' presentation of backup size and number of instruction rollback.  }     
   		\label{fig:case study presentation}     
   	\end{figure}

The Fig ~\ref{fig:case study presentation} illustrates the size of the backups and the number of instruction rollbacks for each method.

\subsection{Backup process} 
If the backup process is complete, the current checkpoint is valid.
If the backup process is incomplete, the valid checkpoint is the lasted checkpoint.


\subsubsection{Case1}
The failure occurred at the location identified by $i$.
If the backup process is complete, so, upon the power restore, both the call-based backup and the exponential incremental call-based backup method could start at the instruction of "$push \ \%rbp$" in $pseudo$. 
If the backup process is incomplete, this backup operation is invalid. 
In the case, both the call-based backup and the exponential incremental call-based backup method could be re-executed the main.

\subsubsection{Case2}
The failure occurred at the location identified by $d$ in the 9th loop.
If the backup process is complete, so, upon the power restore, the call-based backup method could be start at the instruction of "$push\ \%rbp$" in $pseudo$ of 9th loop. 
If the backup process does is incomplete, this backup operation is invalid.
So, upon the power restore, the call-based backup method could start at the instruction of "$push\ \%rbp$" in $pseudo$ of 8th cycle. 
And the exponential incremental call-based backup method could always start at the instruction of "$push\ \%rbp$" in $pseudo$ of 8 loop. 
Because the "$callq$ $pseudo$" of 9th loop is not the checkpoint for the exponential incremental call-based backup method.

\subsubsection{Case3}
The failure occurred at the location identified by $j$ in the main function.
If the backup process is complete, so, upon the power restore, both the call-based backup method and the exponential incremental call-backup method could be start at the instruction of "$push\ \%rbp$" in sum function. 
If the backup process is incomplete, this backup operation is not valid. 
Upon the power restore, both the call-based backup method and the exponential incremental call-based backup method could start at the instruction of "$push\ \%rbp$" in $pseudo$ of 999th loop. 
Because the "$callq$ $pseudo$" of 999th loop is the checkpoint for the two method.

\subsubsection{Case4}
The failure occurred at the location identified by $k$ in the sum function.
if the backup process is complete, so, upon the power restore, both the call-based backup method and the exponential incremental call-based backup method could be start at the instruction of "$push\ \%rbp$" in $pseudo$ function. 
If the backup process is incomplete, this backup operation is not valid. 
Upon the power restore, both the call-based backup method the exponential incremental call-based backup method could start at the instruction of "$push\ \%rbp$" in sum function.

In summary, when there is a power failure during the checkpointing, the program can resume working from the last  complete checkpoint position stored in NVM. If the current checkpoint has been stored in NVM completely, the last checkpoint position is the current call function. 
If it is incomplete, thus the last checkpoint is the latest function call as the checkpoint.

\subsection{Cleanup process}
By default, the cleanup process are based on the correct backup process.

\subsubsection{Case1}
The failure occurred at the location identified by $i$.
Because there are no invalid information in NVM, the cleanup process would not be triggered. 

\subsubsection{Case2}
The failure occurred at the location identified by $d$ in the 9th loop.
For the call-based backup method, regardless of whether the failure occurs before, during, or after cleanup process, when power restore, the call-based backup method could be start at the instruction of "$push\ \%rbp$" in $pseudo$ of 9th loop. 
Because in the 9th backup, the stack frame saved when the pseudo function is called for the 8th backup has been marked as invalid. 
Thus, even if it is not cleaned, it will be removed according to the flag, without affecting the correct recovery of the program. 
And the exponential incremental call-based backup method could always start at the instruction of "$push\ \%rbp$" in $pseudo$ of 8 loop. 
Because the "$callq\ pseudo$" of 9th cycle is not the checkpoint for the exponential incremental call-based backup method, so the cleanup process would not be triggered.

\subsubsection{Case3}
The failure occurred at the location identified by $j$ in the $main$ function.
Upon the power restore, both the call-based backup and the exponential incremental call-based backup could be start at the instruction of "$push\ \%rbp$" in sum function. 

\subsubsection{Case4}
The failure occurred at the location identified by $k$ in the sum function.
Upon the power restore, both the call-based backup method and the exponential incremental call-based backup method could be start at the instruction of "$push\ \%rbp$" in $pseudo$ function. 

In summary, when there is a power failure in the cleanup process, the program can resume working from the last checkpoint position stored in NVM.

Regarding the method of inserting pseudo-functions in multi-layer loops proposed in the section \ref{4.1.5}, there is no specific cases study due to space limited. But it is easy to understand that this method can greatly reduce the write operation to NVM.

\section{Experiments}\label{6}
In this section, we present the experimental evaluation to demonstrate the efficiency of the proposed method.
\subsection{Experiment Setup}
The experiments are conducted with a customized processor simulator. 

We conducted the experiments on a set of benchmarks, such as $adpcm$, $blit$, $bcnt$, $qurt$, and $crc$. 
These benchmarks are from the $powerstone$ benchmark suite \cite{b44}.
The assembly code generated by the GCC compiler for each benchmark is used as input.
As described in section 4, each method is influenced by the setting of various parameters and policies.
Table \ref{table 1} summarizes the parameters and policies of each method.
In Table \ref{table 1}, S=20 indicates that every 20 instructions are statically taken as a backup point, T=20 indicates that a pseudo-function call is inserted every 20 instructions when there are no loops and function calls.
When the times of loop is less than or equal to 10, and there are no function calls inside or outside the loop, we insert a pseudo function call outside the loop as a backup point.
When the times of loop is more than 10, and there are no function calls inside the loop, we insert a pseudo function call outside the loop as a backup point.
In the following experiments, we have added the pseudo function calls to all the pseudo-function call methods with these policies.
After the position of backup points is fixed, 50 failure points are randomly selected in each program to analyze the relationship between the size of backup and the number of rollback instructions at the time of failure.

\begin{table*}[!htp]
\centering
\caption{Parameters value and policies of each backup method}
\label{table 1}
\centering
\begin{tabular}{|c|c|c|}
\hline
backup method&parameters&policy\\
\hline
step-based&S=20 & \\
\hline
\multirow{2}*{pseudo exponential incremental call-based}
& T=20& loop times less than or equal to 10, pseudo call is outside loop \\
\cline{2-3}
& T=20& loop times more than 10, pseudo call is outside loop \\
\hline
\end{tabular}
\end{table*}

\subsection{Evaluation Results}
\begin{figure}[htb] \centering    
    			\includegraphics[width=0.56\linewidth]{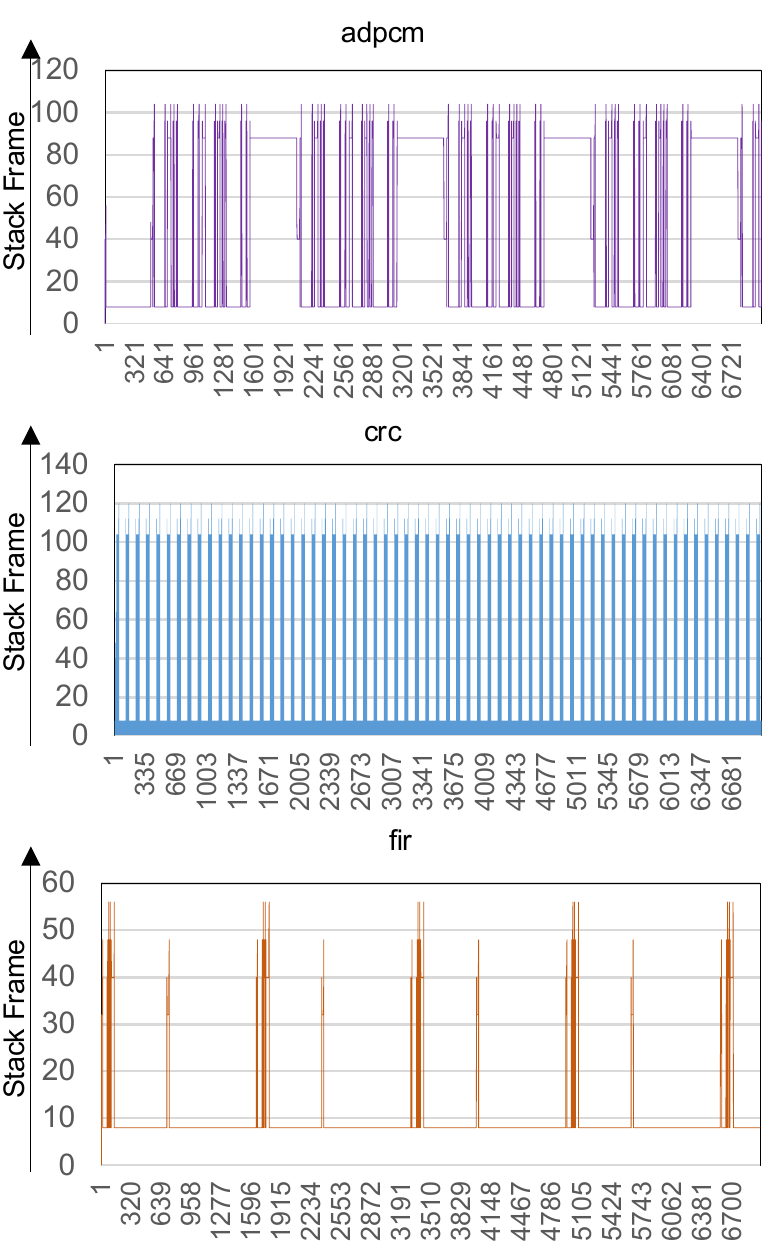}  
    		\caption{Stack frame of each instruction for benchmarks $adpcm$, $crc$, and $fir$.  }     
   		\label{fig:stack frame}     
   	\end{figure}

Fig.~\ref{fig:stack frame} shows the analyzed current stack frame size upon each instruction in representative benchmarks. Here, only 7000 instructions are described, in order to obviously show the fluctuation trajectory. 
The fluctuation verifies the feasibility of using function calls as backup locations.
The proposed method only needs to back up the stack frame (that is, the stack frame of the caller function) when the $call$ instruction is encountered. 
The traditional log-based and step-based method not only need to back up the stack frame when the checkpoint instruction is encountered, but also the stack frame of other caller functions until the $main$ function is backed up.

\begin{figure}[htb] \centering    
    			\includegraphics[width=0.56\linewidth]{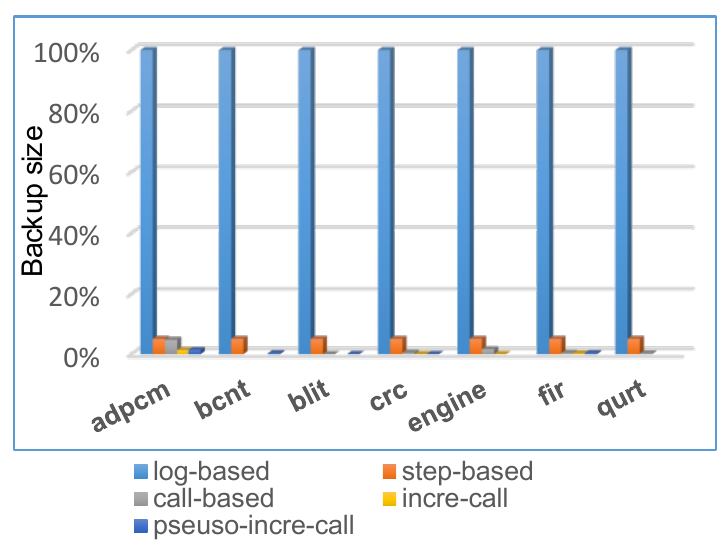}  
    		\caption{The proposed call-based and optimized call-based backup sizes are normalized to the log-based backup size design of each benchmark.  }     
   		\label{fig:backup size}     
   	\end{figure}

Fig.~\ref{fig:backup size} describes step-based, call-based, exponential incremental call-based backup (called incre-call in figure) and pseudo exponential incremental call-based backup (called pseuso-incre-call in figure) the backup content size normalized to the log-based method for tested benchmarks. It shows that the call-based scheme delivers 98.8$\%$ backup content size reduction on average, the exponential incremental call-based scheme delivers 99.6$\%$ backup content size reduction on average, the pseudo exponential incremental call-based scheme delivers 99.6$\%$ backup content size reduction on average, when compared with conventional log-based backup. It also shows that the call-based scheme delivers 80.5$\%$ back up content size reduction on average, the exponential incremental call-based scheme delivers 91.9$\%$ backup content size reduction on average, the pseudo exponential incremental call-based scheme delivers 91.4$\%$ backup content size reduction on average, when compared with conventional step-based backup.

\begin{figure}[htb] \centering    
    			\includegraphics[width=0.56\linewidth]{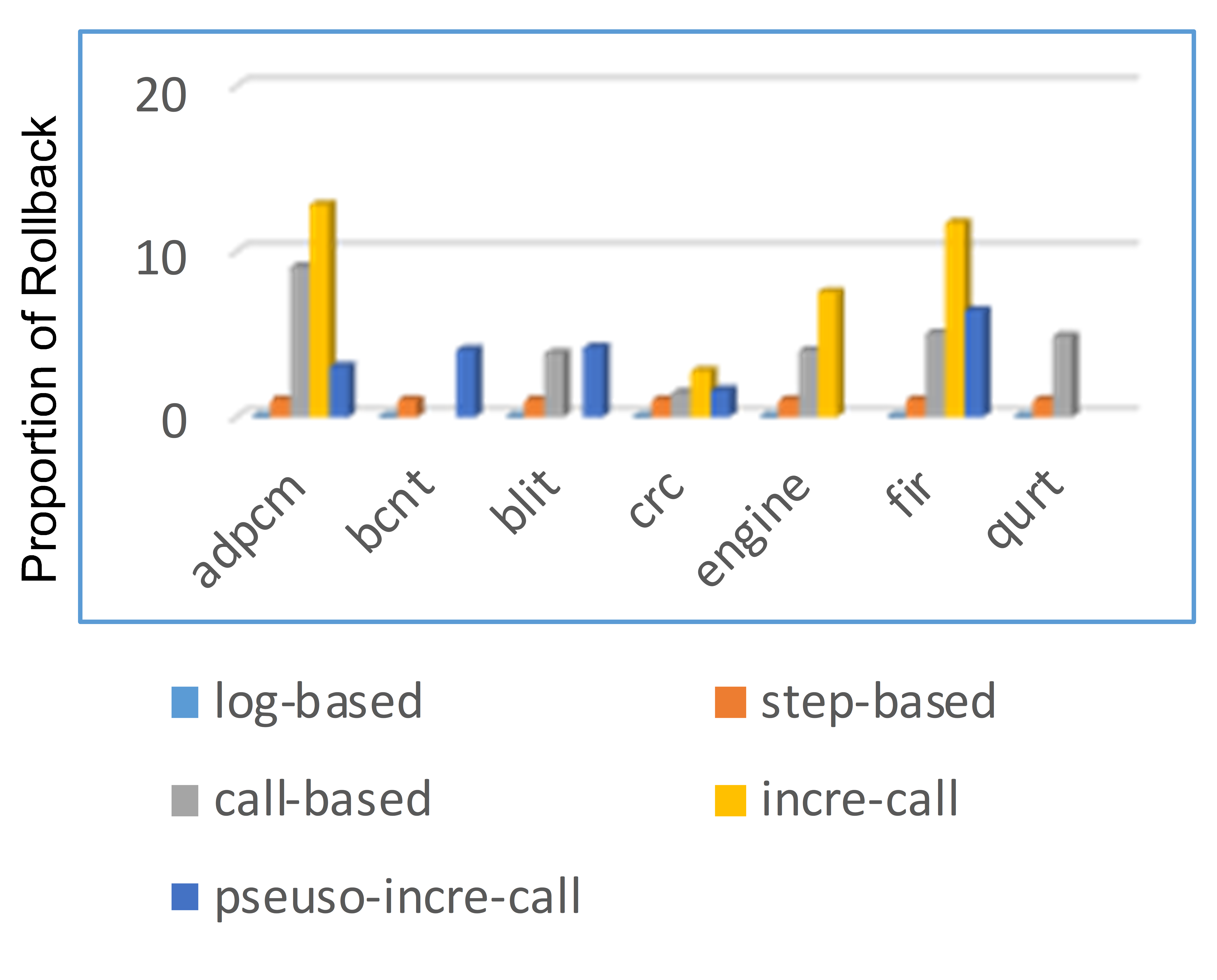}  
    		\caption{The proposed call-based and optimized call-based  backup compared to the step-based number of rollback instruction design of each benchmark.  }     
   		\label{fig:rollback}     
   	\end{figure}

Fig.~\ref{fig:rollback} describes call-based, exponential incremental call-based backup (called incre-call in figure) and pseudo exponential incremental call-based backup (called pseuso-incre-call in figure) the number of rollback instruction compared to the step-based method for tested benchmarks. It shows that the call-based method delivers 5.36X rollback of instruction on average, the exponential incremental call-based method delivers 8.7X rollback of instruction on average, the pseudo exponential incremental call-based method delivers 3.81X rollback of instruction on average. 
Although, the best approach in terms of the number of instruction rollbacks is log-based, the backup content is huge, it is easy to corrupt the NVM, and our method is better tradeoff.


\section{Conclusion}\label{7}

This paper aims to reduce write overheads to NVM during backup while recovering rapidly. 
Different from conventional log-based and step-based method, we propose to use function calls as checkpoints, and through offline analysis, leading to a smaller backup size for NVM. The evaluation shows 99.8$\%$ and 80.5$\%$ NVM backup size reduction on average.
In order to better achieve this, we also proposed inserting pseudo-function calls to increase backup points to reduce recovery costs, and exponential incremental call-based backup methods to reduce backup costs in the loop. 
However, in this paper, we are disable to solution the recursive function. In future work, we will consider the issue.

\vskip3pt

\end{document}